\documentclass[times,11pt,a4paper]{article} 

\makeatletter

\def\@thmcountersep{}
\def\@thmcounterend{:}

\def\mynewtheorem{\@ifstar{\@sthm}{\@Sthm}}

\def\@spnthm#1#2{
  \@ifnextchar[{\@spxnthm{#1}{#2}}{\@spynthm{#1}{#2}}}
\def\@Sthm#1{\@ifnextchar[{\@spothm{#1}}{\@spnthm{#1}}}

\def\@spxnthm#1#2[#3]#4#5{\expandafter\@ifdefinable\csname #1\endcsname
   {\@definecounter{#1}\@addtoreset{#1}{#3}
   \expandafter\xdef\csname the#1\endcsname{\expandafter\noexpand
     \csname the#3\endcsname \noexpand\@thmcountersep \@thmcounter{#1}}
   \expandafter\xdef\csname #1name\endcsname{#2}
   \global\@namedef{#1}{\@spthm{#1}{\csname #1name\endcsname}{#4}{#5}}
                              \global\@namedef{end#1}{\@endtheorem}}}

\def\@spynthm#1#2#3#4{\expandafter\@ifdefinable\csname #1\endcsname
   {\@definecounter{#1}
   \expandafter\xdef\csname the#1\endcsname{\@thmcounter{#1}}
   \expandafter\xdef\csname #1name\endcsname{#2}
   \global\@namedef{#1}{\@spthm{#1}{\csname #1name\endcsname}{#3}{#4}}
                               \global\@namedef{end#1}{\@endtheorem}}}

\def\@spothm#1[#2]#3#4#5{
  \@ifundefined{c@#2}{\@latexerr{No theorem environment `#2' defined}\@eha}
  {\expandafter\@ifdefinable\csname #1\endcsname
  {\global\@namedef{the#1}{\@nameuse{the#2}}
  \expandafter\xdef\csname #1name\endcsname{#3}
  \global\@namedef{#1}{\@spthm{#2}{\csname #1name\endcsname}{#4}{#5}}
  \global\@namedef{end#1}{\@endtheorem}}}}

\def\@spthm#1#2#3#4{\topsep 7\p@ \@plus2\p@ \@minus4\p@
\refstepcounter{#1}
\@ifnextchar[{\@spythm{#1}{#2}{#3}{#4}}{\@spxthm{#1}{#2}{#3}{#4}}}

\def\@spxthm#1#2#3#4{\@spbegintheorem{#2}{\csname the#1\endcsname}{#3}{#4}
                    \ignorespaces}

\def\@spythm#1#2#3#4[#5]{\@spopargbegintheorem{#2}{\csname
       the#1\endcsname}{#5}{#3}{#4}\ignorespaces}

\def\@spbegintheorem#1#2#3#4{\trivlist
                 \item[\hskip\labelsep{#3#1\ #2\@thmcounterend}]#4}

\def\@spopargbegintheorem#1#2#3#4#5{\trivlist
      \item[\hskip\labelsep{#4#1\ #2}]{#4(#3)\@thmcounterend\ }#5}

\def\@sthm#1#2{\@Ynthm{#1}{#2}}

\def\@Ynthm#1#2#3#4{\expandafter\@ifdefinable\csname #1\endcsname
   {\global\@namedef{#1}{\@Thm{\csname #1name\endcsname}{#3}{#4}}
    \expandafter\xdef\csname #1name\endcsname{#2}
    \global\@namedef{end#1}{\@endtheorem}}}

\def\@Thm#1#2#3{\topsep 7\p@ \@plus2\p@ \@minus4\p@
\@ifnextchar[{\@Ythm{#1}{#2}{#3}}{\@Xthm{#1}{#2}{#3}}}

\def\@Xthm#1#2#3{\@Begintheorem{#1}{#2}{#3}\ignorespaces}

\def\@Ythm#1#2#3[#4][#5]{\@Opargbegintheorem{#1}
       {#4}{#2}{#3}{#5}\ignorespaces}

\def\@Begintheorem#1#2#3{#3\trivlist
                           \item[\hskip\labelsep{#2#1\@thmcounterend}]}

\def\@Opargbegintheorem#1#2#3#4#5{#4\trivlist
      \item[\hskip\labelsep{\indent #3#2}]{#3#1 (#5)\@thmcounterend\ }}

\makeatother

\usepackage{indentfirst} 
\usepackage{amsmath} 
\usepackage{latexsym} 
\usepackage{graphics} 
\usepackage{graphicx} 
\usepackage{nextpage}
\usepackage{times} 
\usepackage{amsfonts}
\usepackage{hhline}
\usepackage{url,cite}
\usepackage{tabularx}
\usepackage{epsfig, graphics,subfigure,delarray}
\usepackage{enumitem}
\usepackage{rotating}
\usepackage{verbatim}
\usepackage{multirow}
\usepackage{multicol}
\usepackage{epstopdf}
\usepackage{xcolor}

\mynewtheorem*{kmprprotocol}{Protocol}{\itshape}{\rmfamily}

{\bfseries}{\rmfamily}

\newcommand\mycleardoublepage{\cleartooddpage[\thispagestyle{empty}]}

\def\keywords{\normalfont
    \if@twocolumn
      \small\bfseries\textit{Index Terms}---\,\relax
    \else
      \begin{center}\small\bfseries Index Terms\end{center}\quotation\small
    \fi}
\def\endkeywords{\relax\vspace{0.67ex}
    \par\if@twocolumn\else\endquotation\fi
    \normalsize\normalfont}

\newcommand{\rot}[1]{
\begin{rotate}{45} \textsc{#1} \end{rotate} }

\newcommand\Tstrut{\rule{0pt}{2.2ex}}

\begin{document} 
\pagenumbering{roman}

\begin{figure}[htb!] 
  \begin{center} 
    \includegraphics[width=0.4\textwidth]{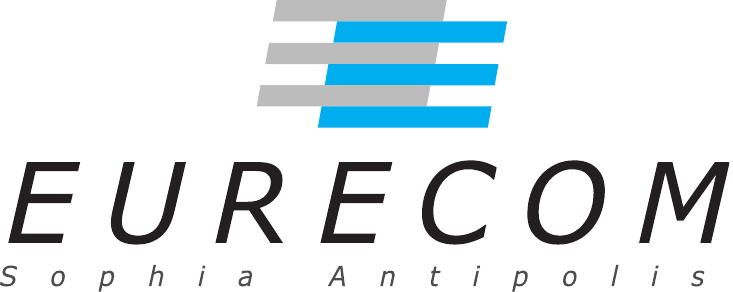} 
  \end{center} 
\end{figure} 

\begin{center} 
  EURECOM\footnotetext[1]{EURECOM's research is partially supported by its industrial members: BMW Group Research and Technology, IABG, Monaco Telecom, Orange, Principauté de Monaco, SAP, SFR, ST Microelectronics, Symantec.}\\ 
  Networking and Security department \\ Campus SophiaTech \\ CS 50193 \\ 06904 Sophia Antipolis cedex 
  \\ FRANCE 
\end{center} 

\vspace{+2cm}

\begin{center} Research Report RR-15-305	 \\  
  \vspace{+5mm}
  \fontsize{14}{12}{\textbf{Taming the Android AppStore: Lightweight Characterization of Android Applications}} \\ 
  \vspace{+5mm}
  April 27\small{$^{\mathrm t \mathrm h}$}, 2015 \\
\end{center} 

\begin{center} 
\vspace{+1cm}
  Luigi Vigneri$^\dagger$, Jaideep Chandrashekar$^\ddagger$, Ioannis Pefkianakis$^\ddagger$ and Olivier Heen$^\ddagger$
  \small
  $^\dagger$\texttt{EURECOM}, \ $^\ddagger$\texttt{Technicolor Research}
\end{center} 

\vspace{4cm} 

\begin{center} 
  Tel : (+33) 4 93 00 81 00 \\ Fax : (+33) 4 93 00 82 00 
\end{center}
\thispagestyle{empty}
\mycleardoublepage
\begin{center}\fontsize{14}{12}{\textbf{Taming the Android AppStore: Lightweight Characterization of Android Applications}} \end{center}

\vspace{+2cm}

\begin{center} 
Luigi Vigneri, Jaideep Chandrashekar, Ioannis Pefkianakis and Olivier Heen 
\end{center}

\vspace{+2cm}

\begin{abstract}
There are over 1.2 million applications on the Google Play store today with a large number of competing applications for any given use or function. This creates challenges for users in selecting the right application. Moreover, some of the applications being of dubious origin, there are no mechanisms for users to understand who the applications are talking to, and to what extent.  In our work, we first develop a lightweight characterization methodology that can automatically extract descriptions of application network behavior, and apply this to a large selection of applications from the Google App Store.  We find several instances of overly aggressive communication with tracking websites, of excessive communication with ad related sites, and of communication with sites previously associated with malware activity. Our results underscore the need for a tool to provide users more visibility into the communication of apps installed on their mobile devices. 
To this end, we develop an Android application to do just this; our application monitors outgoing traffic, associates it with particular applications, 
and then identifies destinations in particular categories that we believe suspicious or else important to reveal to the end-user.
\end{abstract}

\vspace{+2cm}
\thispagestyle{empty}
\mycleardoublepage

\tableofcontents
\pagebreak
\listoffigures
\pagebreak

\pagenumbering{arabic}

\section{Introduction}
\label{sec:intro}

The two dominant mobile application ecosystems today, Apple iOS and Android, reflect very contrasting philosophies. 
In the former, applications are vetted against a defined set of acceptable use behaviors before being released on the store. In contrast, the Android based app stores advocate a more libertarian approach and apply a much looser set of guidelines (mainly focusing on keeping out malicious applications). 
The Google Play Store, the largest and most prevalent Android marketplace today, contains over 1.2 million distinct applications, a number of which are of dubious nature (even if not malicious). The average Android phone user faces a daunting set of questions while installing an application. {\em Which of the many similarly named apps should I install?} {\em Does it report information to online trackers?} {\em Does it have too many ads?} and so on. A search for ``weather'' apps returned well over 30 applications all of which contained {\tt weather} in the name; a large number of these are rated with 4 or more stars (out of 5). Having installed the application, the user has no visibility into who the application is actually communicating with, and whether this complies with the app's intended purpose.

Our goal is to build a system to characterize the network behavior of Android applications. This characterization could inform users about how the application is expected to behave when installed -- useful information when selecting an application. Given our focus on network behavior, 
we are interested in identifying the {\em kinds} of destinations connected to, whether the application connects to a large number of ad sites, how often it talks to online tracking sites, and whether it communicates with sites that have been deemed suspicious. 

In this paper, we first develop a methodology that enables us to characterize a large number of applications quickly. In gaining scale, we necessarily sacrifice a small level of accuracy. Applications connecting to suspicious websites may not actually be sending any private data over the connection. While methods based on taint tracking and static program analysis~\cite{pios,statictaint_2014, Lu:2012, Arzt:2014, Chin:2011, Octeau:2013, Hornyack:2011} enable more accurate characterization of actual data exfiltration, they are intrusive and hard to scale. We view our own work as complementary to such methods, providing a first level characterization that can enable applications to be selected for further, more detailed inspection.  We focus on 3 distinct characteristics, which we believe to be undesirable to end-users, of the destinations being contacted: (i) if they are ad-related, (ii) if they relate to tracking users, or (iii) if the domains have previously been associated with malware or other suspicious activity. 

We then apply this methodology to characterize a large sample of {\em free} applications from the Google Play Store, across different application categories. Our results uncover a great deal of diversity in behavior: some applications connect to almost 2000 different URLs in a few minutes of execution while others generate almost no network traffic. Further, we also identify applications that involve an extensive level of tracking, and those that make an inordinate number of connections to ad related sites.  We also identify application instances that make connections to websites that have previously been associated with malware activity.  These results underscore a crucial shortcoming, and this is a lack of effective tools and mechanisms to audit installed applications and to provide users' greater visibility into application behavior. To this end, we develop a monitoring application (NSA), that identifies particular types of destinations being connected to by installed applications. We have made a version of the application available to reviewers through an anonymous URL.

The rest of the paper is as follows: In Section~\ref{sec:background}, we provide some background on the types of domains that we are interested in characterizing. In Section~\ref{sec:related}, we describe related work in the area and put our own work in perspective. In Section~\ref{sec:dataset}, we elaborate on the process used to collect our dataset and provide some high level summaries. In Sections~\ref{sec:results} and~\ref{sec:characterization}, we present a detailed analysis of the dataset. Section~\ref{sec:framework} contains a high level description of the architecture and operation of the monitoring application we developed, and we conclude in Section~\ref{sec:conc}.

\section{Background}
\label{sec:background}

This work focuses on the network behavior of Android applications. To this end, we are interested in the types of destinations connected to by the application.
 Our methodology, at a high level, consists of extracting network traces from short executions of mobile applications, extracting URL endpoints from the network traces, and categorizing these URLs. 
 To this end, we focus on three distinct types of URL categories which we describe below:

\smallskip
\noindent
{\bf Advertising related sites:} Much of todays online economy is driven by advertising revenue. Web publishers, and mobile app developers, auction space on their websites, or screens, and the ads are delivered by so-called ad networks. Particularly with ({\em free}) mobile applications, advertisements can support app development and support costs, and is a very popular model in the Google Play marketplace. However, most users view advertisements as intrusive and associate negative connotations to them~\cite{tsang2004consumer}.
In order to categorize a URL as being ad-related, we rely on the {\tt EasyList}\footnote{\url{http://easylist.adblockplus.org}} set of filters that are available from AdBlock~\cite{adblock}. This set is able to identify the vast majority of ad serving URLs and can also identify -- based on the URL pattern -- banner ads and  advertisements delivered by other means (javascript, frames, etc.).

\smallskip
\noindent
{\bf Tracking sites:} These provide a mechanism for online web services and third parties to ``follow'' users over multiple sessions, and over different websites. While websites have traditionally used cookies, the app ecosystem uses more direct forms of identification such as UDID or other device identifiers which are made available through the OS APIs. The issue of online tracking has been vigorously debated in the recent past, and privacy advocates argue that it allows for open ended profiling of end-users. Importantly, users are rarely aware of the actual entities that are tracking them, and to what degree and the tracking ecosystem today lacks transparency.

To identify such URL endpoint, we rely on the {\tt EasyPrivacy} set of filters, which are also available through Adblock as an optional subscription. This set covers a large variety of tracking mechanisms (web bugs, tracking code, beacons, etc.).

\smallskip
\noindent
{\bf Suspicious sites:} We use this third category as a catch-all term for any sites that are associated with any form of malware or illicit content, and we do this for the following rationale. Typically, a mobile application is likely to connect to different URLs to carry out the functionality related to the application, apart from the URLs related to ads and tracking. If the application is benign and completely legitimate, it is likely to connect only to destinations that are trusted and safe. However, if the application does make connections to particular websites that, through other channels, have been deemed suspicious or malicious, it is unlikely that the application is completely benign. This likelihood grows larger as the connections to such destinations increases. While not necessarily very accurate -- false positives can happen -- we believe that this factor is one of many that must be considered while vetting the application. Similarly, the mere fact of a connection to a ``malicious'' website may not be evidence of private information being passed on; however, it does arouse suspicion.

In our work, we rely extensively on the VirusTotal meta classification engine~\cite{VirusTotal}, which acts as a front end for a large number of AntiVirus, Spam \& Phishing blacklist, and Malware analysis engines. For each URL identified in the pcap trace, we issue a query to VirusTotal and obtain two types of information for the URL: (i) an aggregate response from all the back-end engines indicating the suspiciousness level of the URL domain, and (ii) a categorization of the URL domain into one of a set of 68 categories. Both of these are obtained from the query responses returned by VirusTotal.

\section{Related Work}
\label{sec:related}

Previous work in mobile application monitoring falls into three broad areas, which we briefly discuss and further contrast them with our work.

{\bf Application profiling:}
Given the lack of insights associated with the Android app store, a number of  studies have focused on profiling mobile apps. 
In~\cite{Wei:2012:profiledroid},  authors describe a multi-layer profiling approach that covers both system and network aspects.
 While capable of obtaining detailed behavioral profiles, the methodology is difficult to scale to a large number of applications, and cannot be implemented as an Android app. 
 In contrast, our (lightweight) approach can characterize a large number of applications. Different from~\cite{Wei:2012:profiledroid}  our study focuses on many aspects of the destinations being connected to by the app. 
 In~\cite{Dai:2013im} the authors describe techniques to fingerprint mobile apps based on their network behavior.
 Our work does not attempt to find application signatures, but  to characterize and compare the network behavior of different apps. 

{\bf Privacy/security auditing:}
Several studies have looked at the issue of identifying privacy leaks in mobile applications. 
TaintDroid~\cite{Enck:2010uw} applies taint tracking mechanisms inside the Android Dalvik VM to identify instruction sequences where a particular input leaves the system. 
The work revealed that roughly half of the tested apps reported the user's location to advertising servers. 
Apart from TaintDroid, there are several systems which seek to identify privacy leaks in iOS \cite{pios}
and Android smartphones \cite{statictaint_2014, Lu:2012, Arzt:2014, Chin:2011, Octeau:2013, Hornyack:2011}, using the taint analysis approach.
While comprehensive, the above approaches often require significant changes in mobile device OS, or phones to be rooted, which limits their applicability. 
Yet another shortcoming with these approaches is that they potentially miss communication to suspicious third party destinations (e.g., trackers) when they do not leak traffic.

Different from taint analysis approaches, SpanDex \cite{SpanDex} extends the Android Dalvik Virtual Machine to ensure that apps do not leak users' passwords.
SpanDex analyzes implicit flows using techniques from symbolic execution to  quantify the amount of information a process control flow reveals about a secret.
ipShield \cite{Chakraborty:2014tr} performs monitoring of every sensor accessed by an app, and uses this information to perform privacy risk assessment.
Both SpanDex and ipShield require modifications in the mobile device OS, while they are not focusing on suspicious destinations.
Finally, in \cite{Grace:2012kv} the authors study the privacy and security risks posed by embedded or in-app advertisement libraries, used in current smartphones.
Our study is much more generic, seeking to identify various characteristics of the mobile apps network behavior.

{\bf Traffic characterization:} Existing work has been focused on analysing smartphone application usage (e.g., geographic coverage, mobility, traffic volume vs. app category) \cite{Xu:2011:IDU}
and traffic characteristics \cite{Maier:2010, Falaki:2010} of mobile devices. None of these studies seek to characterize app communication with third party websites.

\section{Dataset}
\label{sec:dataset}

At a high level, our methodology involves selecting, downloading and executing an application on an {\em unrooted} Android phone, and capturing all the network activity during its execution. The network activity is post-processed to extract all the URLs contacted by the application, which are then categorized and classified using a number of existing online engines. In the rest of this section, we describe the methodology in detail.

\subsection{Application Selection}
The (roughly) 1.2 million applications in the Google App Store today span 25 different categories. 
The choice of category for an application is left to the app developer. As of July 2014, the available categories in the app store are enumerated in 
Table~\ref{tab:categories_play_store}. In each of the categories, the applications can be listed in various orderings -- most popular, most highly rated, newest, etc., and these are accessible with a number of third party APIs. For our characterization, we selected the top 100 {\em most popular} applications and the top 100 {\em newest} applications in each of the categories available, i.e. a total of 5000 applications. However applications can belong to both {\em newest} and {\em most popular} sets, so the actual number of downloaded applications is smaller.
 While not exhaustive, our application set represents a reasonably good sample of the app store (and its categories).
\begin{table}
\centering
{\small 
\begin{tabular}[b]{|c|c|c|}
\hline
Game & News\_and\_magazines & Comics\\
 Libraries\_and\_demo & Communication & Entertainment \\

 Education & Finance & Lifestyle \\

 Books\_and\_reference &  Medical &  Weather\\
 Media\_and\_video &  Music\_and\_audio & Tools \\
 Personalization & Photography &  Productivity\\
 Business &  Health\_and\_fitness & Shopping\\
 Social &  Sports & \\
 Transportation &  Travel\_and\_local & \\
 \hline
\end{tabular}
}
\caption{App Store Categories (July 2014)}
\label{tab:categories_play_store}
\vspace{-0.20in}
\end{table}

\subsection{Application Execution}
From the set of applications selected and downloaded, we filter out all the applications that do not have the {\tt INTERNET} permission property set in the manifest since these applications would not be able to generate any traffic.
Each of the remaining  applications is downloaded and executed on a Samsung Galaxy SIII Mini GT-I8190 smartphone running Android version 4.1.2, which was configured with a VPN client (\emph{OpenVPN for Android}) connected to an external VPN server. All traffic generated on the smartphone transits through the VPN server, where it is captured using {\tt tcpdump}. The manner of network traffic capture differentiates our work from previous work in this field which captured traffic {\em locally} on the smartphone; this is inherently restrictive. The smartphone is connected to a PC, and the app is launched using the {\tt adb} tool on the command line. In order to simulate user interaction with the launched and running application, we use {\tt monkey}, a command-line tool, to generate a series of 10000 user interaction events (screen touches, scroll actions) in two phases, with a short gap of 50 seconds between. This takes care of the situation that some applications have a start-up delay before actuation. A {\tt tcpdump} process is coordinated on the VPN server with each app launch, and thus we obtain a pcap file for each app execution. It is important to point out that while unlikely, there may be other background traffic that co-occurs with application generated traffic. We exercised due diligence in removing all non-essential applications from the smartphone. In addition, we recorded traffic for a 24 hour period without any apps installed on the phone and recorded all traffic generated. We filter out any URLs observed in this trace from each application trace, if found.

\subsection{URL Analysis}
We process each packet capture with {\tt tshark} to extract the complete set of HTTP URLs in the trace. Each of these URLs is classified along three different dimensions, following the previous section (\S\ref{sec:background}). 
Note that we filter out HTTPS traffic (which has been shown to be small in Android apps~\cite{Wei:2012:profiledroid}) 
and only consider HTTP traffic; HTTPS traffic does not expose the headers that we analyze. For each URL extracted, we carry out three different checks as follows:

\begin{enumerate}[leftmargin=0cm,topsep=0cm,itemsep=0cm,itemindent=.5cm,labelwidth=\itemindent,labelsep=0cm,align=left]

\item We check the URL against the set of descriptors in {\tt EasyList} and, if a match is found, we classify the URL as being ad-related, since the connection most likely was made to retrieve an ad-element to display on the smartphone screen.

\item We check the URL against the set of filters contained in {\tt EasyPrivacy} and, if a match is found, we mark the URL as being {\em tracking related}.

\item Finally, we issue a query to the VirusTotal service with the URL as a parameter to obtain a reply that aggregates the findings of all of the backend engines supported by VirusTotal. In addition, we also extract fully qualified domain names from the URL and query VirusTotal for information about these. The results relevant to domain names include things such as the {\tt Webutation} safety score for a domain, and so on.
\end{enumerate}

Finally, our dataset consists of 2146 processed applications (1710 with traffic activity), spanning 25 distinct application categories and which in the aggregate, connect to almost 250k 
unique URLs and across 1985 top level domains. 

\section{Application Destination Characterization}
\label{sec:results}

While there is a considerable body of  work in the area of profiling mobile apps, the focus has been on detecting data leakage, 
or on developing behavioral fingerprints of the applications. There has been relatively less work on characterizing the applications in terms of the network destinations 
they visit, and the nature of these destinations. We focus on analyzing the network end-points in depth and on understanding similarities (or differences) in certain app categories 
in terms of this behavior. We start by presenting some high level statistics of network end-points, across the set of applications analyzed.

\begin{figure}[t]
\subfigure[ \noindent{ Number of URLs per app}.]{
\label{fig:URLs_per_app}
\includegraphics[width=0.48\columnwidth]{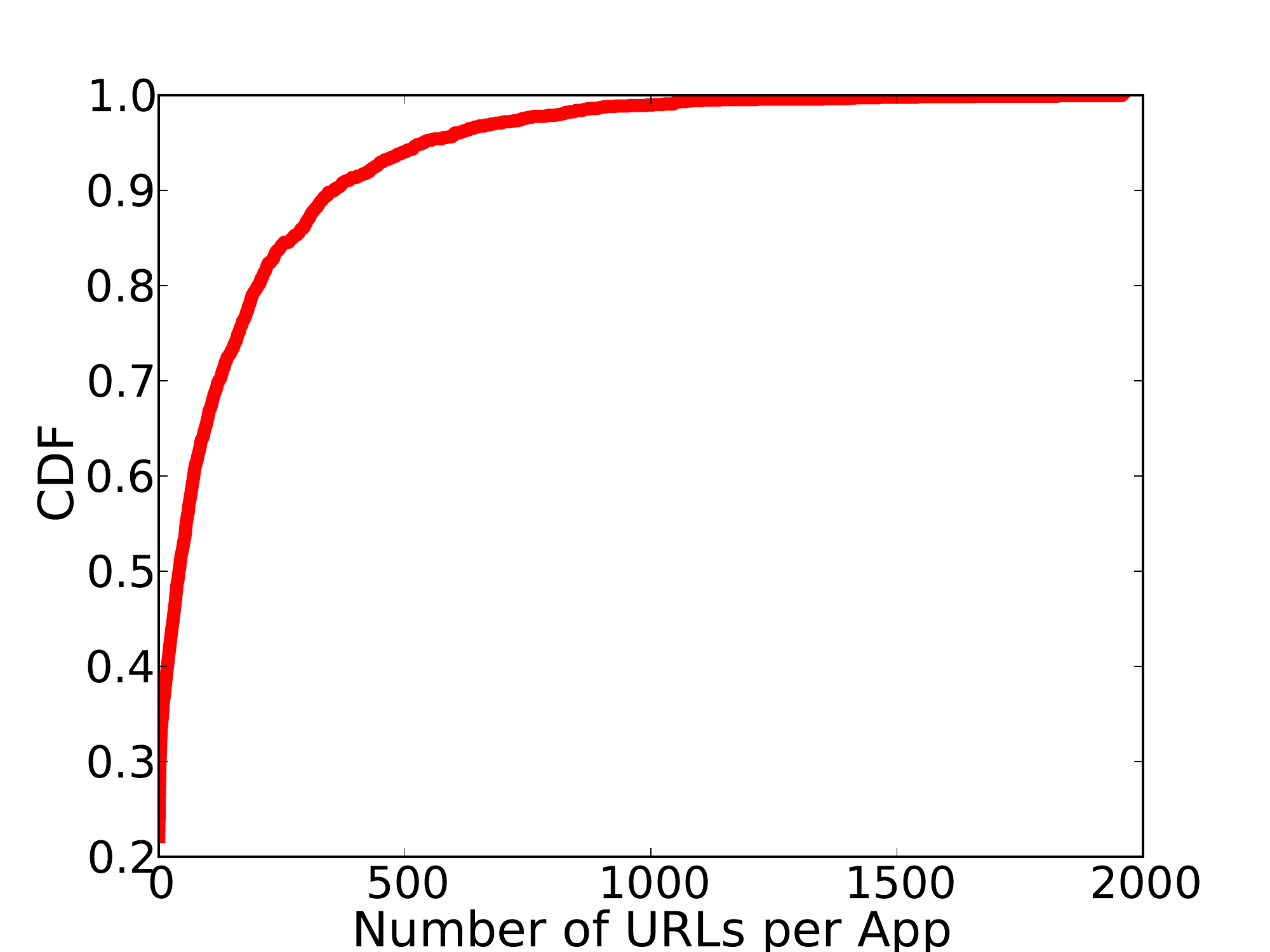} }
\subfigure[ \noindent{Number of domains per app}.]{
\label{fig:domains_per_app}
\includegraphics[width=0.48\columnwidth]{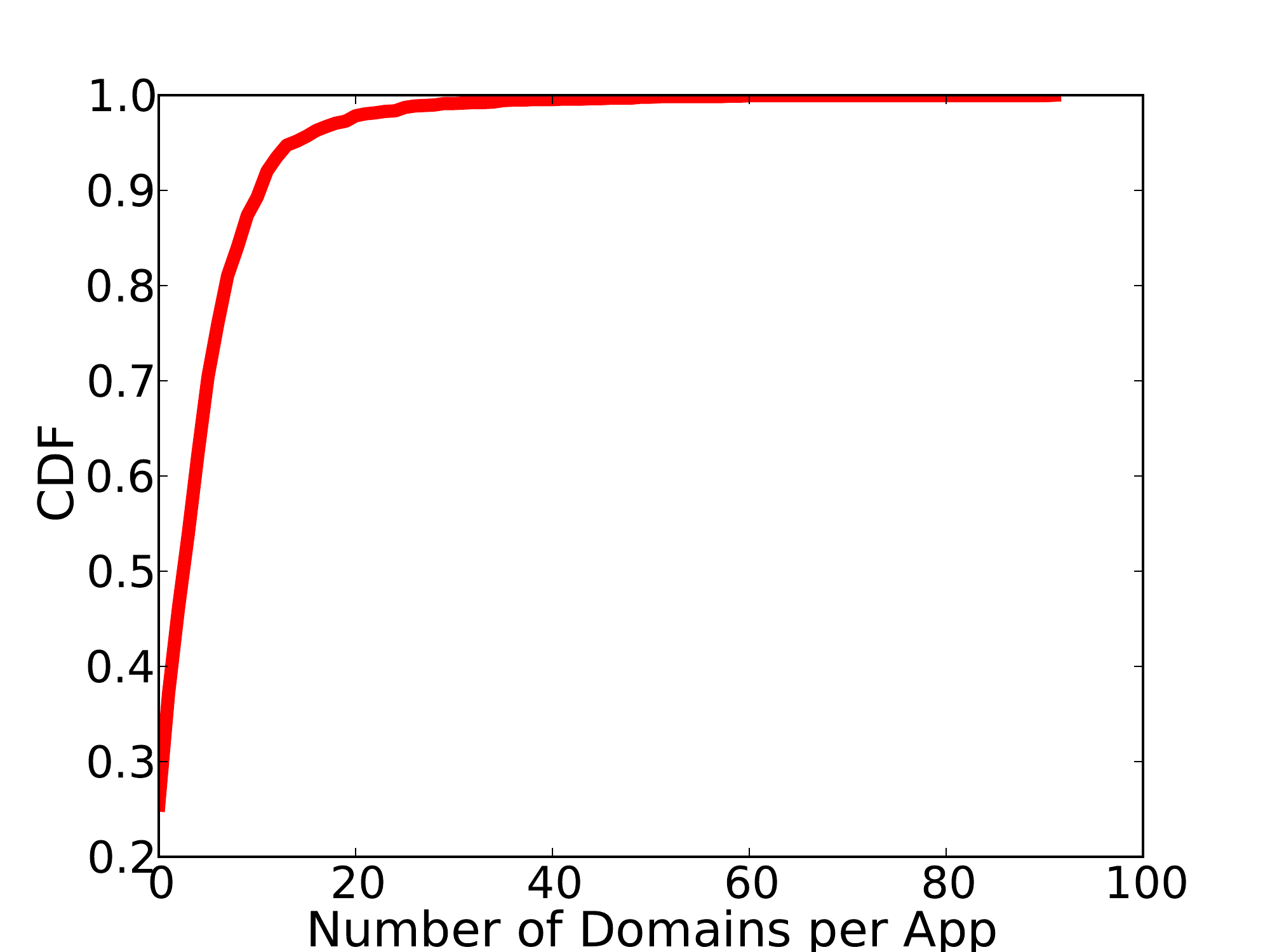} }
\caption{URL and domain counts by application} 
\end{figure}

\textbf{Apps URLs and domains:}
We see a tremendous range in application behavior: a large number of applications generate no traffic at all while some applications generate well in excess of 1000 HTTP requests. 
We find the app {\tt Music Volume EQ} connects to almost {\em 2000 distinct URLs}. Interestingly, Music Volume EQ is a volume slider app, and not an app 
that would really require access to the network. 
 By all accounts, these numbers are large especially considering that our methodology does not support authenticating against user accounts 
(such applications will not progress beyond the login screen). Fig.~\ref{fig:URLs_per_app} shows a distribution of the number of URLs visited by each executed application. 
From the figure, about 10\% of the apps tested connect to more than 500 {\em distinct} URLs (recall that the execution of each application only lasts a few minutes). 
This level of ``chattiness'' significantly impacts resource usage on the mobile device. Interestingly, we still identify apps which do not engage in network activity,
although they declare (in manifest file) that they require network access.

In Table~\ref{tab:apps_by_domain_url}, we enumerate the top 10 applications seen in our dataset, ranked by the number of URLs connected to 
(at least 25 applications connect to more than 1000 URLs during execution). These applications are very diverse, from weather to music and budget.
 This confirms the need to consider broad and varied dataset rather than focusing on specific categories.

\begin{table}
\parbox{0.48\linewidth}{
\centering
{\scriptsize
\begin{tabularx}{0.48\columnwidth}{|X|c|}
\hline
 \textbf{Application Name}  		&\textbf{URLs}\Tstrut\\
\hline
  Music Volume EQ				&1958\Tstrut\\
  signal.booster.conchi...		&1882\\
  Entraînements Quot. FREE		&1827\\
  simulateur laser				&1657\\
  The Weather Channel			&1544\\
  France TV Replay				&1465\\
  Gestion du budget				&1415\\
  Morandini Blog					&1403\\
  FR24 Premium					&1350\\
  cart.tabs.sw					&1275\\
\hline
\end{tabularx}
}
}
\quad
\parbox{.48\linewidth}{
\centering
{\scriptsize
\begin{tabularx}{0.48\columnwidth}{|X|c|}
\hline
 \textbf{Application Name}		&\textbf{TLDs}\Tstrut\\
\hline
 {\em Morandini Blog}			&113\Tstrut\\
 com.issakol12i.myapp			&104\\
 Prof Orientation Test			&103\\
 Exercice Machines Gym			&101\\
 Motor Racing News				& 93\\
 Le Télégramme - Actualité		& 92\\
 app20192.vinebre				& 82\\
 Music Explorer					& 70\\
 PowerAMP Music Player			& 69\\
 {\em Entrainements Quot. FREE}	& 63\\
\hline
\end{tabularx}
}
}
\caption{Top 10 Applications, by URLs (left) and by Top Level Domains (right)}
\label{tab:apps_by_domain_url}
\end{table}

Multiple URLs can correspond to the same domain.
The number of distinct domains, apps connect to, captures the different activities carried out inside the application.  
Across the applications in our dataset, the median number of domains connected to is 4, while some apps connect to more than 100. 
For example, {\tt Morandini Blog}, which is a blog reader application, communicates with 113 distinct domains. 
Interestingly, it connects with 6 {\em different} ad networks, along with a number of analytics and tracking websites. 
In figure~\ref{fig:domains_per_app}, we look across applications and plot the distribution of domains communicated with by each application. 
About half of the apps connect to 4 or fewer domains, and we also see significant variability across applications. 
Roughly 10\% of the apps connect to 20 or more domains over the execution window. 
Table~\ref{tab:apps_by_domain_url} enumerates the top 10 apps ranked by the number of distinct domains connected to.
 Rows marked in italics denote apps that also happen to fall in the top 10 when ranked by the number of URLs communicated with (cf. Table~\ref{tab:apps_by_domain_url}).

\begin{table}
{\scriptsize
\begin{tabularx}{\columnwidth}{ |X|r||X|r| }
\hline
 doubleclick.net        & 0.415& ajax.googleapis.com    & 0.058\Tstrut\\
 google.com             & 0.358& flurry.com             & 0.056\\
 gstatic.com            & 0.354& nend.net               & 0.051\\
 admob.com              & 0.266& xiti.com               & 0.048\\
 googlesyndication.com  & 0.238& facebook.com           & 0.048\\
 google-analytics.com   & 0.172& youtube.com            & 0.041\\
 ggpht.com              & 0.170& scorecardresearch.com  & 0.041\\
 fonts.googleapis.com   & 0.139& inmobi.com             & 0.037\\
 googleusercontent.com  & 0.128& ytimg.com              & 0.034\\
 samsungvideohub.com    & 0.088& twitter.com            & 0.033\\
\hline
\end{tabularx}
}
\caption{Top 20 popular domains (with fraction of applications connecting to them)}
\label{table:domain_popularity}
\end{table}

Looking across applications, Table~\ref{table:domain_popularity} enumerates the 20 most frequently contacted domains, which provides some insights about the nature of communication between the application and website. Unsurprisingly, 9 of the top 10 in this set correspond to various web services run by Google. The most popular domain in the list, {\tt doubleclick.net}, is an advertising platform that tracks end-users, and also serves up advertisments. While {\tt Google.com} is generally considered as the search engine portal, in our traces we found two predominant patterns associated with this particular domain: (i) \url{www.google.com/images/cleardot.gif?zx=<str>}, which correspond to 1x1 tracking pixels, and (ii) \url{www.google.com/ads/user-lists/<id>/?script=<num>&random=<num>}, which seems to indicate some form of user tracking. 

While enumerating the communicating domains can be quite instructive, it does not reveal much about the nature of the communication between app and domain. 
Understanding the {\em type} of domain (or category) of the domain can yield a better sense of this communication. 
 Typically, web domains are set up for well defined functions (e.g., {\tt doubleclick.net} as an ad platform, {\tt google-analytics} as a tracking and analytics service, etc.) 
 and communication between the app and domain is generally consistent with the service offered by the domain.  
 We use the following methodology to identify domain categories. First, we classify each URL as a {\em tracker URL}, {\em ad related} or {\em other}. 
 In the first two cases, the services are obvious, and we consider them independent categories. In the latter case, {\em other}, we extract the fully qualified domain name 
 from the URL and rely on the service provided by {\tt Websense.com} to obtain a characterization (i.e., category) for the domain in question.  
 Specifically, we first examine each URL and extract the fully-qualified domain name (FQDN) embedded in the URL. 
 Then we gather all the Websense categories corresponding to each of the FQDNS that correspond to the same top level domain, 
 and assign the majority class as the category for the top level domain.  
 Going back to the domains listed in Table~\ref{table:domain_popularity}, we find that the most of the 20 domains correspond to the category advertisements. 
 In the rest of this section, we examine various types of destinations based on their domain categorization.

\begin{figure}
 \centering
 \includegraphics[width=0.8\columnwidth]{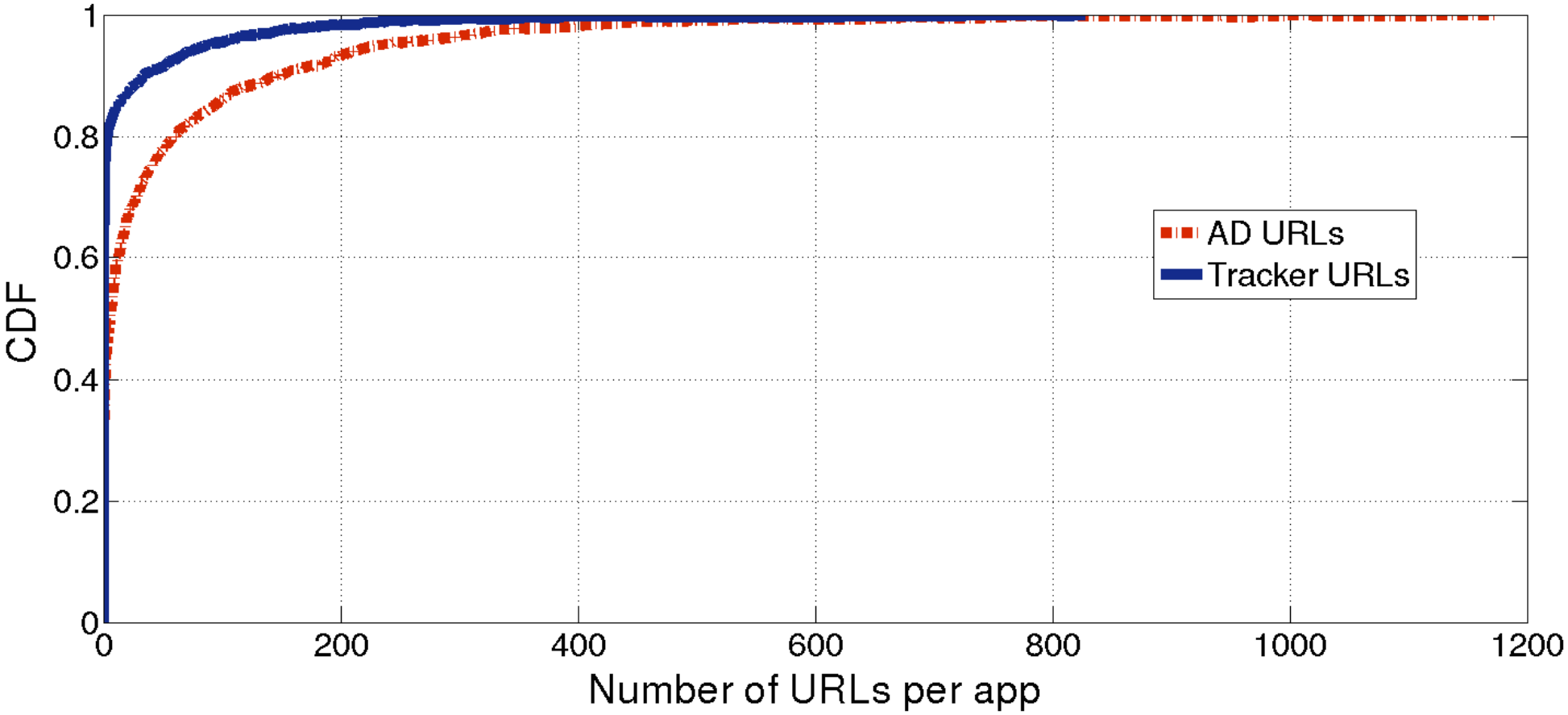}
 \caption{Ad URLs and tracker distributions} 
 \label{fig:tracker_ad_URLs_per_app}
\end{figure}

{\bf Ad related sites:} Recall that we use AdBlock's EasyList subscription to identify advertising related destinations. 
Figure \ref{fig:tracker_ad_URLs_per_app} shows the distribution of the number of ad  URLs visited per app. 
We observe that 33\% of the apps do not communicate with {\em any} ad  destinations. 
On the other hand, we also see apps that connect to a very large number ($>\! 1000$) of ad URLs. Overall, the average number of ad  URLs associated with an application is about 40.  
 Examining the domains of ad URLs, we find that the three most prominent ad related domains are all part of Google, as listed in Table \ref{table:ads_domain_popularity}. 
 Thus, while Google does not directly make any revenue from Android itself (which is openly licensed to manufacturers), it is able to extract revenue from the ads business around the ecosystem. 
 We further investigate ad URLs for particular apps, in Section~\ref{sec:characterization}.

\begin{table}
\begin{center}
\begin{tabular}[b]{|l|c|}
 \hline
  \multicolumn{1}{|c|}{\textbf{Top Level Domain}}    & \textbf{Popularity}\Tstrut\\
 \hline
  doubleclick.net        & 37,153 (50.6\%)\Tstrut\\
  gstatic.com            & 16.532 (22.5\%)\\
  admob.com                & 3603 (4.9\%)\\
  smartadserver.com        & 3411 (4.6\%)\\
  inmobi.com                & 1399 (1.9\%)\\
 \hline
\end{tabular}
\caption{Top 5 ad related Top Level Domains}
\label{table:ads_domain_popularity}
\end{center}
\end{table}

\begin{table}
\begin{center}
{
\begin{tabular}[b]{|l|c|}
 \hline
  \textbf{Top Level Domain}        &    \textbf{Popularity}\Tstrut\\
 \hline
 google-analytics.com    & 11,247 (44.4\%)\Tstrut\\
 xiti.com                & 8174 (32.3\%)\\
 scorecardresearch.com    & 1046 (4.1\%)\\
 estat.com                & 736 (2.9\%)\\
 bluekai.com                & 500 (2.0\%) \\
 \hline
\end{tabular}}
\caption{Top 5 tracking related Top Level Domains}
\label{table:tracker_domain_popularity}
\end{center}
\end{table}

{\bf User tracking related sites:} We now look closely at the URLs in our dataset that correspond to destinations that track end-users and devices, as encoded in AdBlock's EasyPrivacy subscription lists. 
Previous  studies have reported that such practices have largely negative connotations with users~\cite{cmuuser2, cmuuser3}. 
Given this, it is rather surprising that tracking is so widespread, and more important, completely opaque to end-users. 
While the Do Not Track~\cite{dnt} policy has been proposed by consumer advocates and has gained some acceptance, the mechanism is restricted to web browsers, and does not extend to mobile apps in general.

In figure \ref{fig:tracker_ad_URLs_per_app}, we plot the distribution of tracking URLs associated with each application. 
We observe that while the vast majority (73.2\%) of apps do not involve any communication with trackers, a small number of apps do indeed communicate with them. 
The number of tracker URLs per app can be more than 800.
In Table~\ref{table:tracker_domain_popularity}, we enumerate the most popular domains associated with trackers, where popularity is defined as the number of tracker URLs, 
seen across all the apps, associated with a specific domain. In contrast to the results about ad-related destinations, we find the mobile tracking ecosystem to be significantly more fragmented, 
with many more players, even if the dominant player is associated with Google. We further investigate tracker URLs for specific apps, in Section~\ref{sec:characterization}.

\begin{table}
\begin{center}
{\scriptsize
\begin{tabular}[b]{|l || c |}
 \hline
  \textbf{Domain category}    &  \textbf{Popularity}      \\
 \hline
information technology & 453 (22.82\%) \\
uncategorized & 390 (19.65\%) \\
dynamic content & 171 (8.61\%) \\
advertisements & 164 (8.26\%) \\
business and economy   & 110 (5.54\%) \\
news and media & 67 (3.38\%) \\
shopping & 61 (3.07\%) \\
travel & 48 (2.42\%) \\
entertainment & 41 (2.07\%) \\
streaming media & 41 (2.07\%) \\
games & 40 (2.02\%) \\
sports & 37 (1.86\%) \\
search engines and portals & 33 (1.66\%) \\
reference materials & 23 (1.16\%) \\
internet radio and tv & 23 (1.16\%) \\
application and software download & 20 (1.01\%) \\
blogs and personal sites & 17 (0.86\%) \\
vehicles & 17 (0.86\%) \\
social networking & 16 (0.81\%) \\
personal network storage and backup & 15 (0.76\%) \\

\hline
\end{tabular}
}
\caption{Popularity of Domain Categories}
\label{table:domain_cat_pop}
\end{center}
\vspace{-0.20in}
\end{table}

{\bf Other web categories:} We now examine the aggregate set of URLs after having removed those that correspond to the previous two categories. 
Table~\ref{table:domain_cat_pop} enumerates the 20 domain categories with the highest number of domains associated with them. The most popular category, which covers about 22\% of the total domains observed, is denoted {\em Information Technology} and this appears to cover a number of miscellaneous web services. The next two identifiable categories correspond to {\em dynamic content} and {\em advertisements}, and both of these are very likely related to the online advertising ecosystem. 
Note that we see a large count for these even though we filter out those described in {\tt EasyList} previously; the remaining URLs not filtered are likely due to new patterns not in {\em EasyList}  
or perhaps connections to ad related websites that do not involve ad placement inside the mobile application. Apart from these, we see small domain counts across a varied set of categories. 
In the next section, we examine these domain categories in  detail and relate them to the category of the app itself.

{\bf URL {\em badness:}} Finally, we explore an additional  characteristic of the domains being connected to -- ``badness''. 
Recall that VirusTotal aggregates results from a number of engines; these relate to the ``suspiciousness'' of a URL. 
While this term is somewhat ambiguous, the qualitative results can be explained thus: 
the engines used by VirusTotal independently crawl the URLs and catalog the various objects on them. 
URLs that host executable content that is deemed malware-like, are deemed suspicious. 
Note that reliably determining malicious intent is extremely challenging and quite outside the scope of our work. 
For our purpose, we simply quantify whether any engine marked the URL as such, and analyze this across the set of domain categories. 
By {\em suspicion score} for a URL, we denote the fraction of antivirus engines (VirusTotal uses 52 in all) that deem the URL suspicious (or malicious). 
Our result show 94.4\% of the URLs have a (suspicion) score of 0. In the worst case, a URL was deemed suspicious by 3 (of 52) engines.

\textbf{Suspicious domains:} 
For classification of the suspicious domains, we use Webutation engine. Webutation is an open community about Website Reputation. It tests websites against spyware, spam and scams.
Apart from collecting user feedback, Webutation queries various trusted engines like Google Safebrowsing or Norton Antivirus to check for malicious software and other dangerous elements.
Overall, our analysis shows that \emph{a small portion of the domains have been classified by Webutation as suspicious or malicious}.
Specifically, we observe 2.5\% suspicious, 2.9\% malicious, 61\% unsure (not a clear verdict) and  33.6\% safe domains.
Table \ref{table:domain_cat_malicious} further shows the domain categories with the highest fraction of malicious domains.
We observe that the top-3 malicious domain categories are ``sex", ``personals and dating"  and ``ads".  In the following section,
we devise a suspicion metric for mobile apps, and we investigate the most suspicious apps.

\begin{table}
\begin{center}
\begin{tabular}[b]{|c || c |}
 \hline
  \textbf{Domain category}          &  \textbf{Fraction of malicious domains}            \\
 \hline
 sex                                            &  33.33\% \\
personals and dating                 & 20\% \\
ads                                             & 12.8\% \\
business and economy              &  6.6\% \\
reference materials                    &  4.35\% \\
 \hline
\end{tabular}
\caption{Malicious domains based on Webutation engine.
}
\label{table:domain_cat_malicious}
\end{center}
\vspace{-0.30in}
\end{table}

\section{Detailed Apps Characterization}
\label{sec:characterization}

In this section, we focus on individual applications and obtain an understanding of their behavior along the three axes discussed previously. 
Users' value (or are annoyed by) different things -- some user's value privacy (tend to avoid applications with significant tracking), 
other's value security (and wish to avoid applications with suspicious or unreasonable behavior). 
To this end, we study the most prolific applications along these axes and gain some insight into their behavior.

\subsection{Advertising Intensity}
The Internet ecosystem, along with the mobile app marketplace, is largely driven by advertising revenue. 
The vast majority of mobile apps offer their services to the user for free, and are {\em directly} monetized by selling  ``real estate" (smartphone screen or website) 
on which ads are inserted. All of the applications in our dataset are ``free" and we expect that the majority of them {\em will} connect to ad sites.
 This is confirmed in Fig.\ref{fig:tracker_ad_URLs_per_app}, where more than  66\% of the  applications contact ad URLs. 
 Some of the advertising APIs and engines are very aggressive in downloading ads into the mobile app screen. For example {\tt AirPush} is one such infamous service and is so aggressive 
 that the PlayStore lists a number of applications whose sole function is to detect this API and notify the user. Several mobile ad APIs collect detailed device information (OS version, IMEI, location, IP address, etc.), sometimes unknown by users. In general, end users find that ads (esp. display ads which are not targeted based on user interests) degrade the user experience of mobile apps and services. 

Table \ref{tab:top_app_ads} lists the top 10 apps ordered by the number of ad related URLs connected to. All the apps were executed for just a few minutes, and even in this brief interval, we see some apps with a very large number of connections to ad sites. With the exception of two applications -- \emph{Music Volume EQ} and \emph{VidTrim - Video Trimmer} -- none of the others are frequently downloaded (popular), 
and rated positively by users. Note that this information is missing for some of the apps that were removed from the PlayStore soon after we downloaded the APK for testing (and no further information is available).

\begin{table}
{\scriptsize
\begin{tabularx}{\columnwidth}{ |X|c|c|r| }
\hline
 \textbf{Application} & \textbf{Ads} & \textbf{Rate} & \textbf{Downloads} \Tstrut\\
 \hline
 cart.tabs.sw						& 1174	& -		& - \Tstrut\\
 VidTrim - Video Trimmer				& 1065	& 4.2	& 10.000.000\\
 Simulateur laser					& 1019	& 2.1	& 5.000.000\\
 Music Volume EQ						& 999	& 4.2	& 10.000.000	\\
 signal.booster.conchi.amplificador	& 940	& -		& -\\
 com.HillieMelani.VideoEditor		& 720	& -		& -\\
 Football365							& 700	& 3.9	& 10.000\\
 Decibel (Sonometre reactif)			& 671	& 4.8	& 1.000\\
 Nail Art Tutorials 2014				& 630	& 3.7	& 100.000\\
 Veilleuse en Couleurs				& 538	& 3.8	& 10.000\\
\hline
\end{tabularx}
}
\caption{Top 10 apps connecting to ad URLs}
\label{tab:top_app_ads}
\vspace{-0.10in}
\end{table}

\subsection{Tracking Intensity}
We next examine the applications which connect to a large number of tracking services. 
As we observed from Fig. \ref{fig:tracker_ad_URLs_per_app} the vast majority of mobile apps (73.2\%) do not connect to tracker URLs. 
However, the ones that do connect to  trackers tend to connect to a large number. The top 16\% of the apps connect to 100 or more trackers. 
We note an interesting difference between the apps with prolific tracking and those that contact several ad sites. 
The ``high-tracker" apps as shown in Table~\ref{tab:top_app_trackers}, tend to be overall more popular, highly rated and have not been removed fast from Google Play store, compared with the ``high-adveristising" apps.
To help support this argument, we note the incidence of the ``Top Developer Badge" across these sets; Google awards these to developers based on some (opaque) combination of app design, trust and popularity. 
We find 4 of the 10 apps listed in Table~\ref{tab:top_app_trackers} to be associated with this badge, while not a single app listed in Table~\ref{tab:top_app_ads} 
has it\footnote{Only 6\% of the apps in our dataset are by developers with this certification.}. We surmise that this difference is mainly due to how the apps are monetized. 
The ad-driven apps are {\em directly} monetized by the ads they are displaying -- and this  tends to be intrusive to users. 
On the other hand, the apps with a high degree of targeting tend to be more embedded into the online ads ecosystem, 
and generate revenue {\em indirectly} by helping to construct profiles of the smartphone user, which can then be leveraged by them {\em and other apps and services}. 
However this deep integration with the ads ecosystem is difficult for an individual programmer, who has to resort to direct monetization (and which is  likely to push users away).

\begin{table}
{\scriptsize
\begin{tabularx}{\columnwidth}{ |X|c|c|r| }
\hline
 \textbf{Application} & \textbf{Trackers} & \textbf{Rate} & \textbf{Downloads} \Tstrut\\
 \hline
 \emph{Eurosport Player}				& 810	& 3.2	& 500.000 \Tstrut\\
 \emph{RunKeeper}					& 804	& 4.4	& 10.000.000\\
 Gestion du budget					& 725	& 4.3	& 500.000\\
 Logo Quiz							& 301	& 4.5	& 10.000.000	\\
 \emph{Expedia Hotels et Vols}		& 266	& 4.0	& 5.000.000\\
 Vos Droits Quotidien				& 264	& 4.3	& 100\\
 France TV Replay					& 261	& 2.8	& 10.000\\
 \emph{Iron Man 3 Live Wallpaper}	& 250	& 4.0	& 5.000.000\\
 beIN SPORTS							& 236	& 3.6	& 500.000\\
 NipCast								& 235	& 4.7	& 100\\
\hline
\end{tabularx}
}
\caption{Top 10 apps connecting to tracker URLs (italics indicate {\em Top Developer} status)}
\label{tab:top_app_trackers}
\vspace{-0.20in}
\end{table}

\subsection{App Suspiciousness}
Finally we examine the {\em suspiciousness} of each application by leveraging the results from third party engines (as discussed previously). 
Intuitively, we would like an app to be suspicious when contacting many URLs that are tagged as being malicious (or at least, not benign), and {\em more} suspicious when these 
URLs are also spread over several domains. To this end, we define the following metric:
\begin{equation}
 suspicion\_score = \sum_{i\in\textbf{A}}(p_i^\alpha) \cdot d^\beta
 \label{suspicionscore}
\end{equation}
where \textbf{A} is the set of URLs contacted by the  app during its execution, $p_i$ is the (absolute) number of positive (suspicious) signals from VirusTotal for the 
URL $i\in A$ (recall that the result, for each URL query, from VirusTotal is a vector of boolean signals). Finally, $d$ is the distinct number of domains associated with URLs deemed suspicious by VirusTotal. 
The parameters $\alpha\geq 1,\beta\geq 1$ control the ``weights" contributed by the suspicious URLs, or domain cardinality, to the score of the application. 
In our case, we set $\alpha = 3, \beta=1$; while this is somewhat arbitrary, we note that increasing one (or both) parameters simply has the effect of affecting the relative scores between applications. 
In our initial effort, the suspiciousness score was found to be uniformly high across applications and this was attributed to connections made to very popular ad and tracking sites which were flagged by 
VirusTotal- \emph{xiti.com}, \emph{api.airpush.com}, \emph{scorecardresearch.com}, \emph{bluekai.com}, \emph{ad.leadboltapps.net}. 
Whitelisting these popular (also manually verified to be legitimate) domains significantly improved the separation in scores across applications.

Table~\ref{tab:top_suspicious} shows the top 10 applications ordered by the suspiciousness score. Immediately, we see that {\em all} of these apps are associated with a low download count; 
in fact, we found several of these to have been since removed from the PlayStore. Examining these apps in detail, we found several instances where the app name is easily confused with a more well known app. 
We suspect that this ``app name squatting" is deliberately meant to lure customers looking for the bona-fide application. We discuss two examples from our top 10 list in greater detail -- 
\emph{PowerAMP Music Player (BASS)} and \emph{Music Explorer}. The former is named suspiciously similar to the well known {\em PowerAMP} application (which has two developer badge awards). 
In fact, even the package is named to mislead the end-user (\emph{installer.com.maxmpz.audioplayer}, vs. \emph{com.maxmpz.audioplayer} for the bonafide application).
The name ``\emph{Music Explorer}" has been used by several applications in the Google Play Store. Interestingly, the \emph{Music Explorer} app listed as highly suspicious in Table~\ref{tab:top_suspicious}, has been
removed from the Google Store. In conclusion, using simple suspicion metrics (as the one in equation~\ref{suspicionscore}), we can easily blacklist apps, which can potentially harm the mobile device user.

\begin{table}
{\scriptsize
\begin{tabularx}{\columnwidth}{ |X|c|c|r| }
\hline
 \textbf{Application} & \textbf{Score} & \textbf{Rate} & \textbf{Downloads} \Tstrut\\
 \hline
 PowerAMP Music Player			& 13203	& 4.0	& 10.000 \Tstrut\\
 Morandini Blog					& 9758	& 2.4	& 50.000 \\
 Exercice Machines Gym Demo		& 9464	& 4.0	& 10.000 \\
 Motor Racing News				& 7037	& 4.2	& 500 \\
 com.issakol12i.myapp			& 5825	& -		& - \\
 Exercices Quotidien Fessiers*	& 4956	& 4.4	& 1.000.000 \\
 apps.buffalo.kmu.android		& 3480	& -		& - \\
 Music Explorer					& 3366	& 4.4	& 50.000 \\
 biz.pompommanga.readcartoon		& 2920	& -		& - \\
 Notability Basic Guide			& 2820	& 2.3	& 500 \\
\hline
\end{tabularx}
}
\caption{Top 10 apps per suspicion score}
\label{tab:top_suspicious}
\vspace{-0.20in}
\end{table}

\subsection{App Category Behavior}
\label{sec:results_app_vs_domain}

In the previous sections, we focused on the behavior of particular applications. 
We next examine individual app categories with an aim to understand if there is some commonality of behavior among apps of the same category.
Such a characterization is useful in understanding whether a certain application declared by its developer to be in a particular category, 
behaves in a manner consistent with apps of that kind (not suspicious), or whether its behavior shows marked differences compared to others (suspicious).

\begin{figure}
\centering
\includegraphics[width=1.0\columnwidth, height=1.8in]{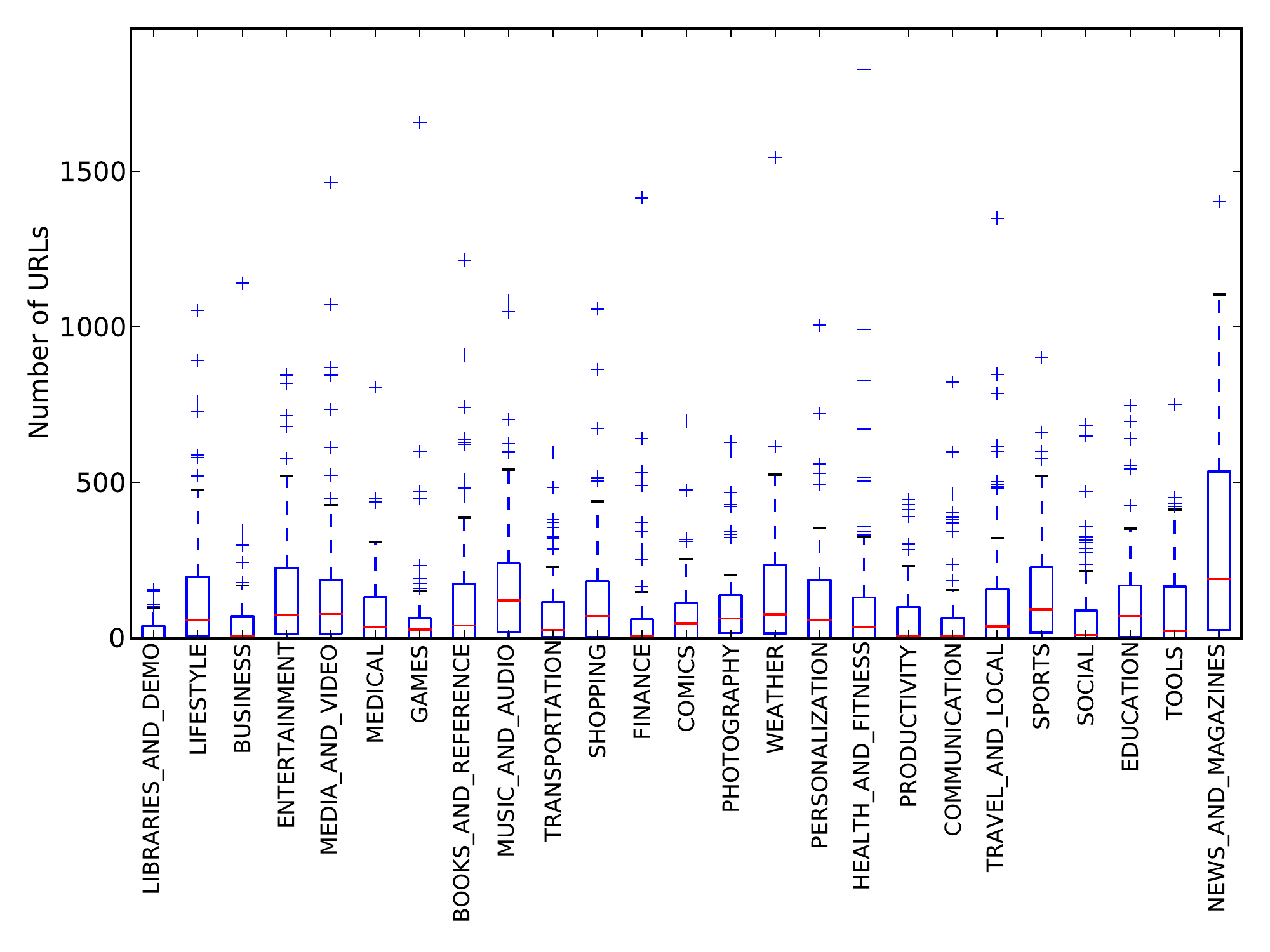}
\caption{Number of URLs for different app categories}
\label{fig:app_category_number_of_URLs}
\vspace{-0.20in}
\end{figure}

We start by looking at the overall number of HTTP connections made by different applications. 
Figure~\ref{fig:app_category_number_of_URLs} shows a boxplot distribution of the number of URLs for each application, organized by category. 
Note that the outliers are indicated by `+' symbols in the graph. As seen in the figure, the one category that does stand out is \emph{NEWS\_AND\_MAGAZINES}, 
where the median number of URLs connected to is about 150.  At the other extreme, the median number of URLs connected to the \emph{LIBRARIES\_AND\_DEMO} category is 1. 
While this seems to suggest that the category of an application strongly influences its chattiness, we do not find statistically significant differences across the remaining categories. 
In fact, the intra- category variation in this value seems to be higher than the inter- category variation. 
Specifically, although the majority of applications in each category connect to very few URLs, a significant number connect to a large number of URLs.

We extend our analysis, by further correlating the app categories with the type of URLs (ads and trackers) and the distinct domains they are connected to.
Overall, we observe apps under \emph{NEWS\_AND\_MAGAZINES} category to connect to higher number of distinct domains, trackers and ad URLs.
However, similar to our previous observation,  the differences among the remaining categories are not statistically significant, while the intra- category variations can be high.
These results show that  current information (such as app category) exposed to users in the Google Store is insufficient, and methods such as ours can provide valuable additional context to users.

 We finally look to whether there are strong associations between app categories and the categories of domains communicated with. 
Table~\ref{table:app_domain_categories} presents the breakdown, in the domains being connected to, across various domain categories for {\em each} application category. 
Each row summarizes a particular set of applications in a category, and each column represents a particular domain category. 
Each table element indicates the fraction of domains related to  a particular app category that belong to a specific domain category. 
For ease of representation, we only present the top 10 categories across all destination domains.  
Overall, we find that the 3 most popular domains relate to (1) advertisements (or domains related to online advertising), 
(2) search engines, and (3) information technology\footnote{IT category is  used as a catch-all term for technology-related websites, when a finer grained characterization is unavailable.}, (respectively). 
These three together account for about 60\% of all the domains connected to. 
Apart from the above dominant domain categories, we observe other popular domain categories for particular app categories.
For example, for \emph{SHOPPING} apps, the third most popular  domain category (with 10.3\% of the domains) is \emph{SHOPPING}.
For \emph{NEWS\_AND\_MAGAZINES} apps, the third most popular  domain category is \emph{SOCIAL}.
However, apart from small exceptions, there is no sufficient correlation between the app category and the domain category, to draw out a systematic fingerprint of the application according to these criteria.
From a user perspective, this pleads for a systematic verification of applications, regardless of their categories.

\section{Application Description}
\label{sec:framework}

The results presented thus far clearly indicate that applications on the Google Play Store often connect to destinations that are not essential for the operation of the app itself. 
Furthermore, much of this communication is completely hidden from users. 
In some cases, the end-user would benefit from an awareness of this activity (e.g., for user-tracking communication). 
Our methodology so far uses of a VPN server to intercept   traffic; this is not practical on phones in active use (the added delay would affect user experience). 

To enable end-users to get this visibility into installed applications, we built an Android application that monitors traffic, using a local proxy, from various installed applications and presents the mobile user additional context about the nature of this traffic and the end points being connected to.

\begin{figure}
\centering
\includegraphics[width=0.6\columnwidth, height=1.1in]{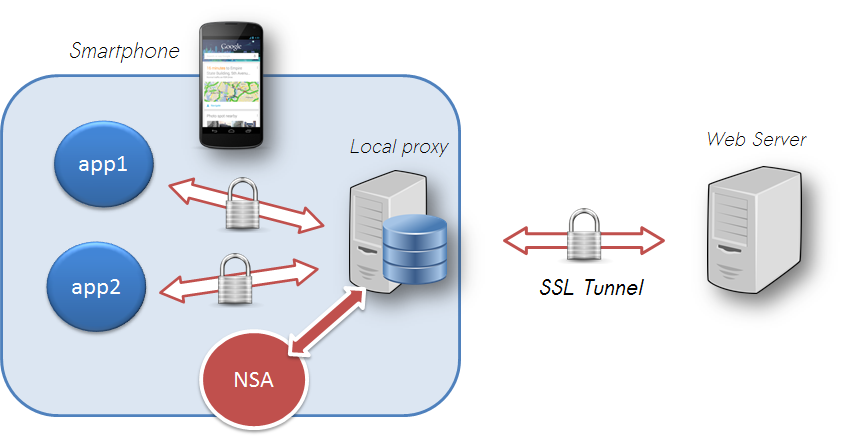}
\caption{Application architecture: HTTP traffic is intercepted by a local proxy service}
\label{fig:app_architecture}
\vspace{+0.20in}
\end{figure}

\begin{table}
\begin{center}
{\tiny

\begin{tabular}{lcccccccccc}

&  \rot{dyn. content}&  \rot{media file dwnld}&  \rot{financial data}&  \rot{shopping}&  \rot{ads}&  \rot{travel}&  \rot{news \& media}&  \rot{IT}&  \rot{search engines}&  \rot{social} \\
\hline

LIBRARIES/DEMO		& 4.9 & 1.6 & & & 26.2 & & 6.6 & 31.2 & 19.5 & \\
LIFESTYLE			& 7.1 & 0.2 & & 1.6 & 23.2 & 1.6 & 4.1 & 24.3 & 15.1 & 4.6 \\
BUSINESS			& 7.9 & & & & 17.0 & & 0.6 & 31.5 & 13.9 & 9.1 \\
ENTERTAINMENT		& 7.8 & 0.5 & 0.3 & 0.8 & 20.1 & & 3.7 & 26.2 & 10.7 & 2.9\\ 
MEDIA/VIDEO			& 6.0 & 0.2 & 0.2 & 0.5 & 26.0 & & 5.2 & 25.1 & 10.3 & 5.1 \\
MEDICAL				& 8.1 & & & & 29.6 & & 5.9 & 27.4 & 9.1 & 3.2 \\
GAMES				& 6.4 & & & 0.8 & 30.1 & & 2.8 & 29.1 & 9.8 & 0.3 \\
BOOKS/REFERENCE		& 6.3 & 0.3 & & 0.6 & 29.8 & & 5.8 & 24.2 & 13.2 & 2.5\\
MUSIC/AUDIO			& 8.8 & 2.6 & & 0.8 & 21.5 & & 3.2 & 24.0 & 9.7 & 5.2 \\
TRANSPORTATION		& 5.1 & & 0.4 & 0.4 & 24.3 & 6.4 & 3.8 & 27.2 & 17.0 & 0.4 \\
SHOPPING			& 8.7 & & & 10.3 & 10.9 & & 2.2 & 25.8 & 8.2 & 9.5 \\
FINANCE				& 6.6 & 1.5 & 8.8 & & 21.2 & & 2.2 & 31.4 & 8.8 & 5.4 \\
COMICS				& 4.4 & & & & 31.9 & & 5.4 & 20.1 & 13.2 & 3.9 \\
PHOTOGRAPHY			& 4.9 & 0.3 & & 0.3 & 30.5 & & 4.9 & 20.1 & 15.2 & 0.9 \\
WEATHER				& 4.8 & 0.3 & & 0.3 & 26.8 & 1.3 & 9.1 & 25.0 & 14.4 & 3.0 \\
PERSONALIZATION		& 8.2 & & & 0.7 & 21.0 & & 6.1 & 28.4 & 17.3 & 0.4 \\
HEALTH/FITNESS		& 5.7 & & & 0.4 & 27.8 & 0.4 & 3.6 & 25.8 & 15.2 & 4.9 \\
PRODUCTIVITY		& 9.6 & & & & 27.0 & & 5.7 & 27.8 & 13.0 & 3.0 \\
COMMUNICATION		& 3.5 & & & 1.3 & 28.0 & & 6.1 & 25.3 & 17.0 & 2.6 \\
TRAVEL/LOCAL		& 2.9 & & 0.3 & 0.5 & 20.3 & 8.8 & 4.0 & 21.3 & 19.5 & 3.5 \\
SPORTS				& 5.4 & 0.3 & & & 18.3 & & 3.0 & 24.3 & 11.4 & 9.9 \\
SOCIAL				& 4.8 & 0.4 & & & 16.7 & 1.1 & 3.7 & 30.1 & 14.5 & 5.2 \\
EDUCATION			& 3.8 & & & 0.6 & 32.6 & & 4.1 & 23.1 & 17.1 & 3.5 \\
TOOLS				& 1.9 & & & & 33.6 & 0.5 & 7.0 & 27.6 & 14.5 & 1.9 \\
NEWS/MAGAZINES		& 3.0 & 0.2 & 0.2 & 0.8 & 19.5 & 0.2 & 8.4 & 30.6 & 10.3 & 11.2 \\
\hline
\end{tabular}
}

\caption{App categories vs. app domain categories. (Data is in percentage (\%))}
\label{table:app_domain_categories}
\end{center}
\vspace{-0.20in}
\end{table}

The high level architecture of our app, which we call NSA (NoSuchApp, in honor of a similarly acronymed monitoring agency) is shown in Fig.~\ref{fig:app_architecture}. We use {\tt SandroProxyLib} to establish a local HTTP (and HTTPS) proxy. By installing its own certificates, the proxy is able to emulate a man-in-the-middle for SSL traffic, which it can then monitor. HTTP(S) requests that are routed through the proxy (which is the case for most applications that use the default network settings) trigger callbacks in our application, which then logs and classifies the HTTP(S) connections. 
While designing NSA, we had to overcome two challenges -- (i) attributing flows to applications correctly, and (ii) classifying the destination URLs without too much overhead. 
To solve the first challenge, NSA periodically polls the system structures in \emph{proc/net/tcp} and \emph{proc/net/udp} (also the structures supporting IPV6), and extracts the mapping between application UID 
and the open sockets (ip address and port). By correlating the socket information with the URLs seen by the proxy, we correctly associate applications with the URLs being generated. 
To solve the second challenge, we batch a number of URLs together before issues queries to various online engines to amortize resource usage (CPU, network). 
Using our application, users can view all the destination end-points associated with a particular application that correspond to one of three categories: 
(i) third party trackers, (ii) suspicious websites\footnote{We currently rely on the characterization provided by the Google Safe Browsing API, but will eventually use the analysis from VirusTotal.}, 
and (iii) destination addresses for which the proxy is bypassed (while there are valid reasons to do so, this is likely also the behavior of malicious applications). 

With this application, our goal is to provide a mechanism for end-users to be aware of the network activity of their installed Android applications. All of the Android users among the authors have been running the application for several weeks, and on one phone (which was previously installed with a rootkit), the application helped identify traffic from applications (that were automatically included with the rootkit) to suspicious destinations (even though the applications were never launched by the user in question). The application is available as an installable package at \url{https://db.tt/Cx8fB5Xz}. 
We plan to make the app publicly downloadable via the Google Play store in the near future.

Looking much further, we can envision a crowdsourced app reputation system driven by NSA where individual users can inspect the traffic being generated by applications tag is as being normal, or else unexpected, or suspicious. Such individual signals could be aggregated at a backend and fed back into the application. This would enable easy blacklisting of applications (and their traffic) based on what other users have observed and reacted to. The exact design of such a crowdsourced system is outside the scope of this paper, and we hope to realize it in future work.

\section{Conclusion}
\label{sec:conc}
The lack of oversight in Android Play Store makes it all too easy for end-users to install applications of dubious origin, or those which silently carry out activity that might not be seen favorably by the user. In this paper, we describe a lightweight characterization methodology that can generate descriptions of application network behavior in an automated manner. The descriptions shed light on the nature of the websites communicated with, focusing on those that may be undesirable to the end-user (ad related, tracking, and malicious).

Using this methodology, we conduct a characterization study of a large number of applications from the Google Play Store. As a general comment we confirm that applications in any domain may carry undesirable activity. This stresses the need for using broad and diversified datasets rather than focussing on specific categories of applications.
In addition, our results reveal several interesting insights: (i) that a significant number of applications, some highly rated, download an excessive number of advertisements which indicate that users may not be as sensitive to advertisements as anecdotally conjectured; (ii) a large number of applications communicate with a multiplicity of online tracking entities, a fact to which users may not be aware; and (iii) we find some applications communicating with websites that have been deemed malicious by malware detection engines. Our results underscore the need for greater transparency in the network interaction of mobile applications on the Android App store(s). To this end, we also describe the design of our own application that provides exactly this service. With our application, end-users are able to understand the different domains the application is communicating with which enables them to make informed decisions about the desirability of the applications they install. 

\bibliographystyle{abbrv}
\bibliography{RR_15_305}

\begin{thebibliography}{10}

\bibitem{adblock}
{AdBlock Plus}.
\newblock \url{https://adblockplus.org/}.

\bibitem{Arzt:2014}
S.~Arzt, S.~Rasthofer, C.~Fritz, E.~Bodden, A.~Bartel, J.~Klein, Y.~Le~Traon,
  D.~Octeau, and P.~McDaniel.
\newblock Flowdroid: Precise context, flow, field, object-sensitive and
  lifecycle-aware taint analysis for android apps.
\newblock In {\em PLDI}, 2014.

\bibitem{Chakraborty:2014tr}
S.~Chakraborty, C.~Shen, K.~R. Raghavan, Y.~Shoukry, M.~Millar, and
  M.~Srivastava.
\newblock {ipShield: A Framework for Enforcing Context-Aware Privacy}.
\newblock In {\em NSDI}, 2014.

\bibitem{Chin:2011}
E.~Chin, A.~P. Felt, K.~Greenwood, and D.~Wagner.
\newblock Analyzing inter-application communication in android.
\newblock In {\em MobiSys}, 2011.

\bibitem{SpanDex}
L.~P. Cox, G.~P., L.~G., P.~V., R.~A., and C.~S. Wu~B.
\newblock Spandex: Secure password tracking for android.
\newblock In {\em USENIX}, 2014.

\bibitem{Dai:2013im}
S.~Dai, A.~Tongaonkar, X.~Wang, A.~Nucci, and D.~Song.
\newblock {NetworkProfiler: Towards automatic fingerprinting of Android apps}.
\newblock In {\em INFOCOM}, 2013.

\bibitem{dnt}
{Do Not Track: Universal Web Tracking Opt Out}.
\newblock \url{http://donottrack.us}.

\bibitem{pios}
M.~Egele, C.~Kruegel, E.~Kirda, and G.~Vigna.
\newblock Pios: Detecting privacy leaks in ios applications.
\newblock In {\em NDSS}, 2011.

\bibitem{Enck:2010uw}
W.~Enck, P.~Gilbert, B.-G. Chun, L.~P. Cox, J.~Jung, P.~McDaniel, and A.~N.
  Sheth.
\newblock {TaintDroid: an information-flow tracking system for realtime privacy
  monitoring on smartphones}.
\newblock In {\em OSDI}, 2010.

\bibitem{Falaki:2010}
H.~Falaki, D.~Lymberopoulos, R.~Mahajan, S.~Kandula, and D.~Estrin.
\newblock A first look at traffic on smartphones.
\newblock In {\em IMC}, 2010.

\bibitem{Grace:2012kv}
M.~C. Grace, W.~Zhou, X.~Jiang, and A.-R. Sadeghi.
\newblock {Unsafe exposure analysis of mobile in-app advertisements}.
\newblock In {\em WISEC}, 2012.

\bibitem{Hornyack:2011}
P.~Hornyack, S.~Han, J.~Jung, S.~Schechter, and D.~Wetherall.
\newblock These aren't the droids you're looking for: Retrofitting android to
  protect data from imperious applications.
\newblock In {\em CCS}, 2011.

\bibitem{cmuuser2}
P.~Leon, B.~Ur, R.~Shay, Y.~Wang, R.~Balebako, and L.~Cranor.
\newblock {Why Johnny Can't Opt Out: a Usability Evaluation of Tools to Limit
  Online Behavioral Advertising}.
\newblock In {\em ACM CHI}, 2012.

\bibitem{statictaint_2014}
L.~Li, A.~Bartel, J.~Klein, Y.~L. Traon, S.~Arzt, S.~Rasthofer, E.~Bodden,
  D.~Octeau, and P.~McDaniel.
\newblock I know what leaked in your pocket: Uncovering privacy leaks on
  android apps with static taint analysis.
\newblock Technical Report ISBN 978-2-87971-129-4, University of Luxembourg,
  April 2014.

\bibitem{Lu:2012}
L.~Lu, Z.~Li, Z.~Wu, W.~Lee, and G.~Jiang.
\newblock Chex: statically vetting android apps for component hijacking
  vulnerabilities.
\newblock In {\em CCS}, 2012.

\bibitem{Maier:2010}
G.~Maier, F.~Schneider, and A.~Feldmann.
\newblock A first look at mobile hand-held device traffic.
\newblock In {\em PAM}, 2010.

\bibitem{Octeau:2013}
D.~Octeau, P.~McDaniel, S.~Jha, A.~Bartel, E.~Bodden, J.~Klein, and
  Y.~Le~Traon.
\newblock Effective inter-component communication mapping in android with
  epicc: An essential step towards holistic security analysis.
\newblock In {\em Usenix Security}, 2013.

\bibitem{tsang2004consumer}
M.~M. Tsang, S.-C. Ho, and T.-P. Liang.
\newblock Consumer attitudes toward mobile advertising: An empirical study.
\newblock {\em International Journal of Electronic Commerce}, 8(3), 2004.

\bibitem{cmuuser3}
B.~Ur, P.~Leon, L.~Cranor, R.~Shay, and Y.~Wang.
\newblock {Smart, Useful, Scary, Creepy: Perceptions of Online Behavioral
  Advertising}.
\newblock In {\em SOUP}, 2012.

\bibitem{VirusTotal}
{VirusTotal Meta Engine}.
\newblock \url{https://www.virustotal.com/}.

\bibitem{Wei:2012:profiledroid}
X.~Wei, L.~Gomez, I.~Neamtiu, and M.~Faloutsos.
\newblock Profiledroid: Multi-layer profiling of android applications.
\newblock In {\em Mobicom}. ACM, 2012.

\bibitem{Xu:2011:IDU}
Q.~Xu, J.~Erman, A.~Gerber, Z.~Mao, J.~Pang, and S.~Venkataraman.
\newblock Identifying diverse usage behaviors of smartphone apps.
\newblock In {\em IMC}, 2011.

\end{thebibliography}
\end{document}